%Paper: 9108007
%From: leclair@lnssun3.tn.cornell.edu (Andre LeClair)
%Date: Tue, 20 Aug 91 15:52:20 EDT

%%%%%%%%%%%%%%%%%%%%%%%%%%%%%%%%%%%%%%%%%%%%%%%%%%%%%%%%%%%%%%%%%
%%%%%%  `Infinite Quantum Group Symmetry of Fields in Massive
%%%%%%      2D Quantum Field Theory'   by A. LeClair and F. Smirnov  
%%%%%%%%%%%%%%%%%%%%%%%%%%%%%%%%%%%%%%%%%%%%%%%%%%%%%%%%%%%%%%%%%%%    
\input harvmac

\def\form{F. A. Smirnov, in {\it Introduction to Quantum Groups and 
and Integrable Massive Models of Quantum Field Theory}, 
 Nankai Lectures in Mathematical Physics, Mo-Lin Ge and 
Bao-Heng Zhao (eds.), World Scientific, Singapore (1990)\semi
A. N. Kirillov and F. A. Smirnov, Phys. Lett. 198B (1987) 506; 
Zap. Nauch. Semin. LOMI 164 (1987) 80.}

%
%
%
%%%%%%%%%%%%%%%%%%%%%%%%%%%%%%%%%%%%%%%%%%%%%%%%%%%%%%%%%%%%%%%%%%%%%%

%

\def\Smirii{F. A. Smirnov, Commun. Math. Phys. 132 (1990) 415.}

\def\Luschi{M. L\"uscher, Nucl. Phys. B135 (1978) 1.}

\def\KaTh{M. Karowski and H. J. Thun, Nucl. Phys. B190 (1981) 61.}

\def\Yang{D. Bernard, {\it Hidden Yangians in 2D Massive Current Algebras,} 
to appear in Commun. Math. Phys. }

\def\ABL{C. Ahn, D. Bernard, and A. LeClair,  
Nucl. Phys. B346 (1990) 409. }

%%%%%%%%%%%%%%%%%%%%%%%%%%%%%%%%%%%%%%%%%%%%%%%%%%%%%%%%%%%%%%%%%%%%%%%%%

\def\KniZam{V.G. Knizhnik and A.B. Zamolodchikov, Nucl. Phys. B297
(1984) 83.} 
\def\GepW{D. Gepner and E. Witten, Nucl. Phys. B278 (1986) 493.}

%%%%%%%%%%%%%%%%%%%%%%%%%%%%%%%%%%%%%%%%%%%%%%%%%%%%%%%%%%%%%%%%%%%%%%%%%

%

%
%

%
%

%

%
%

%

%

%
%

%

%

%

%
%
%
%
  
%

%
%

%
%
 
%
%

%
%
 
%
%

%
%

%
%
 
%
%

%
%
\def\BPZ{A. A. Belavin, A. M. Polyakov, and A. B. Zamolodchikov, 
Nucl. Phys. B241 (1984) 333.}

\def\Drinfeld{V. G. Drinfel'd, Sov. Math. Dokl. 32 (1985) 254;
Sov. Math. Dokl. 36 (1988) 212. }
\def\Jimbo{M. Jimbo, Lett. Math. Phys. 10 (1985) 63; Lett. Math. Phys. 11 
(1986) 247; Commun. Math. Phys. 102 (1986) 537.}

\def\Zamolodchikovi{A. B. Zamolodchikov and Al. B. Zamolodchikov, Annals
Phys. 120 (1979) 253.}
%

%

%

%

%

%

%

%

%

%

%

%

%

%

%

%

%

%

%

%

%

%

%

%

%

%

%

%%%%%%%%%%%%%%%%%%%%%%%%%%%%%%%%%%%%%%%%%%%%%%%%%%%%%%%%%%%%%%%%%%%%%%%
%%%%%%%%%%%%%%%%%%%%%% END REFERENCES %%%%%%%%%%%%%%%%%%%%%%%%%%%%%%%%%%
%%%%%%%%%%%%%%%%%%%%%%%%%%%%%%%%%%%%%%%%%%%%%%%%%%%%%%%%%%%%%%%%%%%%%%

%%%%%%%%%%%%%%%%%%%%%%%%%%%%%%%%%%%%%%%%%%%%%%%%%%%%%%%%%%%%%%%
%
%		DEFINITIONS FOR TEX
%
%%%%%%%%%%%%%%%%%%%%%%%%%%%%%%%%%%%%%%%%%%%%%%%%%%%%%%%%%%%%%%%
%

%
%\def\e{\'e}
%\def\ee{\`e}
%%%%%%%%%%%%%%%%%%%DEFINITIONS%%%%%%%%%%%%%%%%%%%%%%%%%%%%%%%%%
%

\def\bra#1{{\langle #1 |  }}

\def\bar{\overline}

\def\*{\star}
\def\[{\left[}
\def\]{\right]}
\def\({\left(}		
\def\){\right)}

%
%%%%%%%%%%%%%%%%%%%%%%%%%%%%%%%%%%%%%%%%%%%%%%%%%%%%%%%%%%%%%%%
%
\def\frac#1#2{{#1 \over #2}}
\def\inv#1{{1 \over #1}}

\def\d{\partial}
\def\der#1{{\partial \over \partial #1}}

\def\ket#1{ | #1 \rangle}
\def\rvac{\hbox{$\vert 0\rangle$}}
\def\lvac{\hbox{$\langle 0 \vert $}}
\def\2pi{\hbox{$2\pi i$}}

\def\dsl{\raise.15ex\hbox{/}\kern-.57em\partial}
\def\Dsl{\,\raise.15ex\hbox{/}\mkern-.13.5mu D}
%
%%%%%%%%%%%%%%%%%%%%GREEK LETTERS%%%%%%%%%%%%%%%%%%%%%%%%%%%%%%
%

\def\be{\beta}
\def\al{\alpha}
\def\ep{\epsilon}
	
\def\de{\delta}		\def\De{\Delta}

%
%%%%%%%%%%%%%%%%%%%CALIGRAPHIC LETTERS%%%%%%%%%%%%%%%%%%%%%%%%%
%
\def\CA{{\cal A}}		\def\CC{{\cal C}}
\def\CD{{\cal D}}		
	\def\CH{{\cal H}}	
\def\CJ{{\cal J}}		
	\def\CN{{\cal N}}	\def\CO{{\cal O}}
		\def\CR{{\cal R}}
	\def\CT{{\cal T}}	
		
\def\CY{{\cal Y}}	\def\CZ{{\cal Z}}

\def\rvac{\hbox{$\vert 0\rangle$}}
\def\lvac{\hbox{$\langle 0 \vert $}}

\def\2pi{\hbox{$2\pi i$}}

\def\dsl{\raise.15ex\hbox{/}\kern-.57em\partial}
\def\Dsl{\,\raise.15ex\hbox{/}\mkern-.13.5mu D}
%
%%%%%%%%%%%%%%%%%%%%GREEK LETTERS%%%%%%%%%%%%%%%%%%%%%%%%%%%%%%
%
%%%%%%%%%%%%%%% MATH CHARACTERS %%%%%%%%%%%%%%%%%%%%%%%%%%%%
%
\font\numbers=cmss12
%\font\numbers=cmu10 scaled\magstep1
\font\upright=cmu10 scaled\magstep1
\def\stroke{\vrule height8pt width0.4pt depth-0.1pt}
\def\topfleck{\vrule height8pt width0.5pt depth-5.9pt}
\def\botfleck{\vrule height2pt width0.5pt depth0.1pt}
\def\Zmath{\vcenter{\hbox{\numbers\rlap{\rlap{Z}\kern 0.8pt\topfleck}\kern 2.2pt
                   \rlap Z\kern 6pt\botfleck\kern 1pt}}}
\def\Qmath{\vcenter{\hbox{\upright\rlap{\rlap{Q}\kern
                   3.8pt\stroke}\phantom{Q}}}}
\def\Nmath{\vcenter{\hbox{\upright\rlap{I}\kern 1.7pt N}}}
\def\Cmath{\vcenter{\hbox{\upright\rlap{\rlap{C}\kern
                   3.8pt\stroke}\phantom{C}}}}
\def\Rmath{\vcenter{\hbox{\upright\rlap{I}\kern 1.7pt R}}}
\def\Z{\ifmmode\Zmath\else$\Zmath$\fi}
\def\Q{\ifmmode\Qmath\else$\Qmath$\fi}
\def\N{\ifmmode\Nmath\else$\Nmath$\fi}
\def\C{\ifmmode\Cmath\else$\Cmath$\fi}
\def\R{\ifmmode\Rmath\else$\Rmath$\fi}
%%%%%%%%%%%%%%%%%%%%%%%%%%%%%%%%%%%%%%%%%%%%%%%%%%%%%%%%%%%%%%%%%
 %%%%%%%%%%%%%%%%%% END OF DEFINITIONS %%%%%%%%%%%%%%%%%%%%%%
 %%%%%%%%%%%%%%%%%%%%%%%%%%%%%%%%%%%%%%%%%%%%%%%%%

\Title{CLNS 91-1056}
{\vbox{\centerline{Infinite Quantum Group Symmetry of Fields  } 
\centerline{ in Massive 2D Quantum Field Theory} }}

\bigskip
\bigskip

\centerline{Andr\' e LeClair}
\medskip\centerline{Newman Laboratory}
\centerline{Cornell University}
\centerline{Ithaca, NY  14853}
\bigskip
\centerline{and}
\bigskip
\centerline{F. A. Smirnov}
\medskip 
\centerline{Leningrad Branch of the Steklov Mathematical Institute}
\centerline{Fontanka 27, Leningrad 191011, USSR} 

\vskip .3in

Starting from a given S-matrix of an integrable quantum field 
theory in $1+1$ dimensions, and knowledge of its on-shell
quantum group symmetries, we describe how to extend the symmetry
to the space of fields. This is accomplished by introducing
an adjoint action of the symmetry generators on fields, and 
specifying the form factors of descendents.  The braiding
relations of quantum field multiplets is shown to be given by
the universal $\CR$-matrix.  We develop
in some detail the case of infinite
dimensional Yangian symmetry.
We show that 
the quantum double of the Yangian is a Hopf algebra deformation
of a level zero Kac-Moody algebra that preserves its finite
dimensional Lie subalgebra.  The 
fields form infinite dimensional Verma-module representations; 
in particular the energy-momentum tensor and isotopic current
are in the same multiplet.

\Date{3/91}
%\draft
%
%

%
%
%
%
%sample reference
%
%on an operator formulation of the superstring\ref\a{\sdual} . 
%
%sample equations

%\eqn\one{
%\V 123 A\rangle_1 \vb_2 \vc_3 = 
%\langle h_1 \left[ V_A (0)\right] h_2 \left[ V_B (0) \right] 
%h_3 \left[ V_C (0) \right] \rangle . }
%
%
%
%\eqnn\two
%$$\eqalignno{
% h_1 (z ) &= z \cr
 %h_2 (z ) &= { 1\over {1-z } }&\two\cr
% h_3 (z ) &= { {z -1}\over {z} } .\cr}$$
%

%\newsec{Introduction }

%\centerline{Acknowlegments} 
%\figures
%\fig{1}{bla} 
%\listrefs
%\end

%
%
%
%
%
%
\def\del#1#2{{ \de^{#1#2} }}
\def\Phib{{\bf \Phi}}
\def\m#1#2#3{m_{#1#2}^{#3} } 
\def\u#1#2#3{\mu^{#1#2}_{#3} }
\def\ad#1{ {\rm ad}_{\displaystyle {#1}} } 
\def\adp#1{ {\rm ad}' _{\displaystyle {#1}} } 
\def\qo#1{Q_0^{#1}}
\def\qp#1{{Q_1^{#1}}}
\def\qm#1{Q_{-1}^{#1}} 
\def\BLnlc{D. Bernard and A. LeClair, {\it Quantum Group 
Symmetries and Non-Local Currents in 2D QFT,} 
to appear in 
Commun. Math. Phys;   Phys. Lett. B247 (1990) 309.} 

\newsec{Introduction} 

The study of quantum field theory in $1+1$ spacetime dimensions 
can provide important insight into the structure of quantum 
field theory in general.  Of particular interest are the 
exactly solvable massive models, which are integrable 
Hamiltonian systems.  In many cases, the exact on-shell
S-matrices are known. 
(For some well-known examples see
\ref\rzz{\Zamolodchikovi} .)  The exact form factors for some fields
in many of the models have also been computed \ref\rsmiri{\form} 
. 
Despite the integrability of the models, many  of their important
properties, most notably their correlation functions, remain
unknown. 

More recently it has been recognized that the massive integrable 
quantum field theories are to a large extent characterized
by quantum group symmetries, which can serve to define the
theory non-perturbatively 
\ref\rsmirii{\Smirii}\ref\rdenis{\Yang}
\ref\rbl{\BLnlc} . 
The quantum group symmetry algebras of interest are infinite 
dimensional, and 
non-abelian.  Explicit currents that generate a 
q-deformation of affine Kac-Moody algebras \ref\rdrin{\Drinfeld}
\ref\rjimbo{\Jimbo}\ were constructed in the sine-Gordon theory
and its generalization to imaginary coupling affine Toda theory 
\rbl , and shown to completely characterize the S-matrices.  
At a specific value of the coupling constant where the theory 
is invariant with respect to a finite dimensional Lie algebra 
$G$, these theories are the Gross-Neveu models, and there the infinite
dimensional symmetry becomes the Yangian of Drinfel'd, which contains 
$G$ as a subalgebra.  
The non-local conserved currents that generate this symmetry 
were actually constructed some time ago by L\"uscher\ref\rlus{\Luschi} . 
However the connection 
of these currents  to Yangians was only recently recognized by
Bernard \rdenis . 

The existence of these infinite dimensional symmetries leads to the 
idea that these models may be solvable in the spirit that the 
conformal field theories are solved \ref\rcon{\BPZ} . 
In this methodology the emphasis is on how the fields realize  
infinite dimensional Verma-module representations of the 
symmetry algebra.

In this work, assuming  the exact S-matrix  
 and its
on-shell symmetry for a general model are given, 
we describe how the symmetry can be 
extended off-shell to the space of fields. 
We begin by defining an adjoint action of the symmetry
algebra on fields, which  generalizes the notion of commutator
one encounters for ordinary symmetries, to an arbitrary quantum
group symmetry.  Form factors of `descendent' fields are shown
to be explicitly computable from the form factors of 
`ancestor' fields. The braiding of fields in the multiplets 
obtained by adjoint action is given by the universal $\CR$-matrix
of the quantum group.  
A result of this kind was also found for the finite dimensional
quantum group symmetry of conformal field theory by 
Gomez and Sierra\ref\sigo{C. G\'omez and G. Sierra, 
Phys. Lett. 240B, (1990), 149; 
`The Quantum Symmetry of Rational Conformal Field Theory', to
appear in Nucl. Phys. B.}. 

The example we develop in some detail, which displays all of the 
interesting issues involved, is characterized by $sl(2)$ Yangian
symmetry.  
As mentioned above, these models can be formulated 
as the $G$-invariant Gross-Neveu model \ref\rgn{\KaTh} , where 
$G$ is here   taken to be $sl(2)$.  
This model can also be formulated in a way more suitable for our
discussion
as  a massive current-current
perturbation of the $G$-invariant Wess-Zumino-Witten-Novikov
(WZWN) model at level $k= 1$, with the action
\eqn\wzw{
S ~=~ S_{G}^{ {\rm level}~1}  +  g
\int d^2 x ~ J^a (x) \bar{J} ^a (x) , } 
where  
$J^a , \bar{J}^a $ are the level 1 Kac-Moody currents of the 
WZWN conformal field theory $S_G$.  
For this model, we are able to construct infinite dimensional 
multiplets through adjoint action.  The derivative of the 
$sl(2)$ current is actually a descendent of the energy-momentum
tensor.  We construct the appropriate braiding matrix by
studying the quantum double for the Yangian.  These braiding
matrices are infinite dimensional. 

It is interesting to compare some of the features of the massive verses 
the conformal theory. 
The conformal model described by 
$S_{G}$ can be 
completely solved using its  Kac-Moody symmetry \ref\rkz{\KniZam}
\ref\rgepw{\GepW}, 
where the Kac-Moody generators are modes of the currents, 
$J^a_n = \oint \frac{dz}{2\pi i} ~z^n \> J^a (z)$, $n\in \CZ$. 
 These generators satisfy the algebra 
\eqn\km{
\[ J^a_n , J^b_m \] = f^{abc} \> J^c_{n+m} ~ + ~ \inv{2} k \, n 
\> \delta^{ab} \> 
\delta_{n, -m} ,  } 
The full symmetry of the conformal model also includes 
and additional antiholomorphic Kac-Moody symmetry generated
by $\bar{J}^a_n = \oint \frac{d \bar{z} } {2\pi i } \>  \bar{z}^n 
\> \bar{J}^a (\bar {z} )$. 
As we will describe, the Yangian algebra of the massive theory 
is a deformation of a Kac-Moody algebra that preserves its finite
dimensional Lie subalgebra $G$. 
However the Yangian symmetry does not appear to be related in any
simple way to the Kac-Moody symmetry of the conformal model.  
In fact, setting the deformation parameter of the Yangian to zero
one recovers a level 0 Kac-Moody algebra.  This zero value for the level
is a general feature of the infinite dimensional algebras relevant for the
massive theories.  Indeed the theories with q-deformed affine Kac-Moody
symmetry, which the Yangian invariant theories are a limiting case of, also
have zero level \rbl .  

This zero value of the level has a physical explanation. Level zero
Kac-Moody algebras (sometimes called loop algebras) possess finite 
dimensional representations.  In conformal field theory, since there is 
a one-to-one correspondence  between fields and states, a non-zero
level is required for the infinite dimensional representations corresponding
to the fields.  In contrast, for massive models, if the symmetry algebra
commutes with the particle number operator, then the symmetry algebra
must possess finite dimensional representations on the space of particle 
states, since the number of states at fixed particle number is finite. 
Thus, assuming that at least some of the representations of the  Kac-Moody
algebra can be deformed into representations of the Yangian, the level 
must be zero. 
However, in distinction with conformal field theory, since there is not 
this  correspondence between states and fields in massive theory, 
the fields of the theory may still form distinct  infinite dimensional 
representations.  

This paper is organized as follows. In the first section we review
some aspects of the theory of quantum groups that we will need. 
The general, model independent formalism is presented in section 3. 
The example of Yangian symmetry is treated in section 4. 

\newsec{Quantum Groups} 

In this section we review some general aspects of the theory
of quantum groups, following mainly the presentation of 
Drinfel'd \ref\rdrinb{V. G. Drinfel'd, `Quantum Groups', 
{\it Proceedings of the International Congress of 
Mathematicians}, Berkeley, CA, 1986.}\ 
and Faddeev, Reshetikhin, and Takhtajan 
\ref\rfadd{L. Faddeev, N.  Reshetikhin, and L. Takhtajan, 
in {\it Braid Group, Knot Theory and Statistical Mechanics}, 
C . N. Yang and M. L. Ge (ed.),  World Scientific, 1989. } . 

\def\am{{ $\CA$ }} 
\def\a{\CA}
\def\ot{\otimes} 

Quantum groups are non-trivial examples of Hopf algebras. 
Let $\CA$ denote such  
a Hopf algebra.  It is equipped with a multiplication map
$m: \CA \otimes \CA \to \CA$, a comultiplication 
$\Delta : \a \to  \a \ot \a$, antipode $s: \a \to \a$, and 
counit $\varepsilon : \a \to \CC$, where $\CC$ is the complex
numbers. 
\def\vep{\varepsilon} 
\def\De{\Delta} 
We suppose that \am contains the unit element 1, with 
$\Delta(1) = 1\ot 1, s(1) = \vep (1) = 1$.  These operations
have the following properties: 
\eqna\Ii
$$\eqalignno{
m(a\ot 1) &= m(1\ot a) = a &\Ii {a} \cr 
m(m\ot id ) &= m (id \ot m ) &\Ii {b} \cr
(\Delta \ot id ) \Delta &= (id \ot \Delta )\Delta &\Ii {c} \cr
\De (a) \De (b) &= \De (ab)  &\Ii {d} \cr 
m (s\ot id ) \De (a) &= m (id \ot s ) \De (a) = \vep (a) \cdot 1 
&\Ii {e} \cr 
s(ab) &= s(b) s(a) &\Ii {f} \cr
\De s &= (s\ot s ) P  \De  &\Ii {g} \cr
(\vep \ot id ) \De &= (id \ot \vep ) \De = id  &\Ii {h} \cr
\vep (ab) &= \vep (a) \vep (b) , &\Ii {i} \cr }$$
for  $a,b \in \a$, and $P$ is the permutation operator
$P(a\ot b) = b\ot a $. 
Eq. \Ii{a} ~ is the definition of the unit element, \Ii{b,c}\ are 
the associativity and coassociativity of \am , \Ii{d} defines 
$\De$ to be a homomorphism of \am to $\a \ot \a$, and \Ii{e-i}\ 
are the defining properties of the counit and antipode. 

Let $\{ e_a \}$ denote a linear basis for \am .  The above operations
can be formulated by means of structure constants: 
\eqna\Iii
$$\eqalignno{
e_a \>  e_b &= \m abc \; e_c &\Iii {a} \cr
\De (e_a ) &= \u bca \; e_b \ot e_c &\Iii {b} \cr
s(e_a ) &= s_a^b \; e_b  &\Iii {c} \cr
\vep (e_a ) &= \vep_a , &\Iii {d} \cr }$$
where $\m abc , \u bca , s_a^b $, and $\vep_a $ are constants
in $\CC$. 
We also define constants $\vep^a$ such that 
\eqn\Iiib{ 1= \vep^a e_a .} 

The properties \Ii{}\ are easily expressed as consistency
conditions on the structure constants:
\eqna\Iiii
$$\eqalignno{
\m abc \> \vep^a &= \m bac \> \vep^a = \delta_b^c &\Iiii {a} \cr
\m abc \> \m deb &= \m adb \> \m bec &\Iiii {b} \cr
\u bca \> \u deb &= \u dba \> \u ecb &\Iiii {c} \cr
\u ija \> \u klb \> \m ikc \> \m jld &= \m abi \> \u cdi &\Iiii {d} \cr 
\m ijb \> s^j_k \> \u ika &= \m jib \> s^j_k \> \u kia = \vep_a \vep^b 
&\Iiii {e} \cr
\m abi \> s_i^c &= s_b^j \> s_a^k \> \m jkc &\Iiii {f} \cr
\u abi \> s^i_c &= s^a_j \> s^b_k \> \u kjc &\Iiii {g} \cr
\u bca \> \vep_c &= \u cba \> \vep_c = \delta^b_a &\Iiii {h} \cr
\m abc \> \vep_c &= \vep_a \vep_b &\Iiii {i} \cr
}$$

For the applications we are interested in, we need to introduce 
the concept of the quantum double.  Let $\a^*$ denote an 
algebra dual to \am in the quantum double sense, with 
basis $\{ e^a \} $.  The elements of the dual basis are 
defined to satisfy the following relations: 
\eqna\Iiv
$$\eqalignno{
e^a e^b &= \u abc \> e^c &\Iiv {a} \cr
\De (e^a ) &= \m bca \> e^c \ot e^b &\Iiv {b} \cr
s(e^a ) &= (s^{-1} ) ^a_b \> e^b  &\Iiv {c} \cr
\vep (e^a ) &= \vep^a  &\Iiv {d} \cr
\vep_a e^a &= 1 . &\Iiv{e} \cr }$$
Due to the properties of the structure constants \Iiii{},
one has a Hopf algebra structure on $\a^*$. 

The quantum double is a Hopf algebra structure on the space 
$\a \ot \a^* $. The relations between the elements 
$e_a$ and the dual elements $e^a $ are defined as follows. 
Define the permuted comultiplication $\De'$: 
\eqn\Iv{
\De' (e_a ) = \u bca \> e_c \ot e_b , } 
and skew antipode $s'$: 
\eqn\Ivi{
m(s' \ot 1 ) \De ' (a)  = \vep_a \cdot 1 . } 
The  universal 
$\CR$-matrix is an element of $\a \ot \a^*$ defined by 
\eqn\Ivii{
\CR = \sum_a e_a \ot e^a , }
satisfying 
\eqn\Iviii{
\CR ~ \De (e_a ) = \De' (e_a ) \CR ,}  
for all $e_a$. 
The equation \Iviii\ implies the following relation between 
elements of \am and $\a^*$: 
\eqn\Iix{
\u bca \> \m ibd \> e^i \, e_c = \u cba \> \m bid \> e_c \, e^i  . } 
Using the property \Ivi\ of $s'$, and \Iiii{b,c}, 
the above equation can be written as 
\eqna\Ix
$$\eqalignno{ 
e_a \, e^b &= \mu^{kcl}_a \> m^b_{idk} \> {s'}^i_l ~e^d \, e_c &\Ix {a} \cr
e^b \, e_a &= \mu_a^{lck} \> m_{kdi}^b \> {s'}^i_l ~e_c \, e^d  &\Ix {b} \cr
}$$
where 
\eqn\Ixb{ \mu^{lck}_a = \u lia \> \u cki ; ~~~~~~ 
m^b_{kdm} = \m kdi \> \m imb . } 

\def\r{\CR} 

The universal $\CR$-matrix constructed in this fashion satisfies the 
Yang-Baxter equation
\eqn\Ixi{
\CR_{12} \CR_{13} \CR_{23} = \CR_{23} \CR_{13} \CR_{12} . } 
One also has 
\eqn\Ixib{ 
\r^{-1} = \sum_a s(e_a ) \ot e^a . } 

When \am has an invariant bilinear form, $\a^*$ can often be identified with
\am , and $\r$ becomes an element of $\a \ot \a$.  However in 
some important examples, such as the Yangian, this is not the case. 

\newsec{General Theory} 

Let us suppose we are given the exact factorizable 
S-matrix of an integrable quantum field theory.  As 
usual, the energy-momentum of an on-shell particle is 
parameterized by the rapidity $\be$ 
\eqn\IIi{
p_0 (\be ) = m \cosh (\be )  ,~~~~~p_1 (\be ) = m \sinh (\be ) . } 
In general, the one-particle states have isotopic degrees of 
freedom taking values in some vector space $V$. A complete
infinite dimensional set of states is provided by the multiparticle
basis 
\eqn\IIii{
\ket B = \ket {\be_n , \be_{n-1} , ...., \be_1 } .}
The above state is a vector in $V^{\ot n}$;  the isotopic indices
are not explicitly displayed.  In-states have 
$\be_n >\be_{n-1} > \cdot\cdot\cdot >\be_1$, whereas out-states have 
$\be_1 >\be_2 >\cdot\cdot\cdot > \be_n$. 

The two-particle to two-particle S-matrix is an operator
\eqn\IIiii{
S_{12} (\be_1 -\be_2 ) ~:~~V\ot V \to V \ot V . } 
The S-matrix $S_{12} (\be_1 -\be_2 ) $ satisfies the usual 
constraints of the Yang-Baxter equations, unitarity and 
crossing symmetry \rzz :
\eqna\IIiv
$$\eqalignno{
S_{12} (\be_1 -\be_2 ) S_{13} (\be_1 -\be_3 ) S_{23} (\be_2 -\be_3 ) 
&= S_{23} (\be_2 -\be_3 ) S_{13} (\be_1 -\be_3 ) S_{12} (\be_1 -\be_2 ) 
&\IIiv {a} \cr
S_{12} (\be ) S_{21} (-\be ) &=1  &\IIiv {b} \cr
S_{12} (i\pi -\be ) &= c_2 S_{21} (\be ) c_2 , &\IIiv {c} \cr
}$$
where $c$ is the matrix of charge conjugation, and the subscripts refer 
to the space where the operator acts.  The multiparticle S-matrix 
is factorized into 2-particle S-matrices: 
\eqn\IIivb{
S_{i,i+1} (\be_i -\be_{i+1} ) ~ 
\ket {\be_n , \be_{n-1} , .., \be_{i+1} , \be_i , .., \be_1 } 
= \ket {\be_n , \be_{n-1} , .., \be_i , \be_{i+1}  , .., \be_1 } 
.} 

We further suppose that the S-matrix is invariant under some 
symmetry algebra \am . The algebra \am is 
assumed to be a quantum group with the structures
outlined in section 2. The  space of multiparticle 
states $\CH$ forms an infinite
dimensional representation $\rho_{\CH}$ of the algebra \am . 
It is assumed that the generators of \am commute with the particle
number operator.  This implies that the representation $\rho_{\CH}$ 
is necessarily reducible into finite dimensional representations 
of fixed particle number. 
(More specifically we have finite dimensional representations at fixed
rapidities, i.e. we do not count the additional multiplicity involved
in varying the rapidity.)
\def\rh{\rho_{\CH}} 
Given the representation $\rh$ of \am on one-particle states, 
which is in general rapidity dependent, the representation on
multiparticle states is provided by the comultiplication 
$\De$.  The on-shell symmetry of the S-matrix is the 
statement
\eqn\IIv{
S_{12} (\be_1 -\be_2 ) ~ \rho_V \ot \rho_V \( \De (e_a ) \) 
= \rho_V \ot \rho_V \( \De ' (e_a ) \) ~ S_{12} (\be_1 - \be_2 ) . } 

For the models of interest, \IIv\ completely characterizes the 
S-matrix. 
Comparing eq. \IIv\ with the defining equation \Iviii\ for the 
universal $\CR$-matrix it is evident that the S-matrix 
is a specialization of the $\CR$-matrix to the finite dimensional
rapidity dependent representation $\rho_V$ of \am 
(up to overall scalar factors $s_0 (\be)$ required for unitarity and 
crossing symmetry): 
\eqn\IIvb{ 
S_{12} (\be ) = s_0 (\be ) ~ \rho_V \ot \rho_V \( \CR \) . } 

It is of primary interest to extend the above on-shell symmetries 
to the space of fields.  This problem is more complex than for the 
conformal field theories since there is no one-to-one correspondence
between fields and states here.  
The fields in the theory are completely defined by specifying 
their matrix elements in the space of states $\CH$, i.e. their
form factors.  Let us begin by reviewing some basic properties 
of form factors in integrable quantum field theory \rsmiri . 
The form factors $f( \be_1 , ...., \be_n )$ of a field 
$\phi (x)$ are defined as the particular matrix elements  
\eqn\IIvi{
f(\be_1 , ..., \be_n ) = \bra 0 \> \phi (0 ) \> \ket {\be_n ,\be_{n-1} , ..., \be _1 } 
. } 
The form factor in \IIvi\ is a map from $V^{\ot n}$ to a function
of $n$ rapidity variables;  it is thus a vector in the dual 
space $\( V^{\ot n} \) ^* $.  

The form factors satisfy the following axiomatic properties: 
\eqn\IIvii{
S_{i,i+1} (\be_i - \be_{i+1} ) f(\be_1 , ..., \be_i , \be_{i+1} , 
..., \be_n ) 
= f(\be_1, ..., \be_{i+1} , \be_i , ..., \be_n ) , } 
 
and for local fields 
\eqn\IIviii{
f(\be_1, ..., \be_{n-1} , \be_n + 2\pi i ) 
= f(\be_n , \be_1, \be_2 , ..., \be_{n-1} ) . } 
The form factors also satisfy a third axiom that relates residues
of $n$-particle form factors to $n-1$-particle form factors, 
but we will not need it here. 
The general matrix elements follow from the functions \IIvi\ 
by crossing symmetry
\eqn\IIix{
\bra {\al_1 , ..., \al_m } \> \phi (0) \> \ket {\be_n ,..., \be_1} 
= c_1 \cdot\cdot c_m \> f (\al_m -i\pi , ..., \al_1 -i\pi , 
\be_1 , ..., \be_n ) . }
(The equation \IIix\ is correct as it stands for the set
of rapidities $\{ \al_1 ,..., \al_m \} $ disconnected from 
the set $\{ \be_1 , ..., \be_n \}$.  When some of the 
rapidities $\al$ coincide with $\be$, there are some additional 
$\delta$-function terms in \IIix . ) 
Finally the $x$ dependence of the matrix elements can be trivially
restored due to translational invariance
\eqnn\IIx
$$\eqalignno{
\bra {\al_1 , ..., \al_m } \> \phi (x) \>  \ket {\be_n ,..., \be_1} 
&= \exp \( i x_\mu \( \sum_{i=1}^m p_\mu (\al_i ) 
- \sum_{i=1}^n p_\mu (\be_i ) \)  \) \cr  
&~~~~~\cdot\bra {\al_1 , ..., \al_m } \> \phi (0) \>
\ket {\be_n ,..., \be_1} . 
&\IIx\cr}$$  

For the more familiar symmetries of quantum field theory, such
as global Lie-algebra invariance, the symmetry is realized on the 
space of fields through the commutator with the global conserved
charges.  The Jacobi identity ensures that fields related by 
this adjoint action fall into finite dimensional representations
of the symmetry algebra.  It is this feature that we now 
generalize to an arbitrary quantum group \am . We define
two adjoint actions on the space of fields:\foot{For arbitrary
Hopf algebras there exists a standard definition of the adjoint action
of the algebra on itself which is analagous to the adjoint action
on fields defined 
below \ref\reshpriv{N. Yu. Reshetikhin, private communication.}.} 
\eqna\IIxi
$$\eqalignno{
\ad {e_a } \( \phi (x) \) &\equiv \u ija \> s(e_i ) \, \phi (x) \, e_j  
&\IIxi {a} \cr 
\adp {e_a } \( \phi (x) \) &\equiv \u ija \> s' (e_j ) \, \phi (x) \, e_i 
~. &\IIxi {b} \cr }$$
The significance of the two different adjoint actions will be explained
shortly. 
The adjoint actions as we have defined them are primarily characterized
as being homomorphisms 
\eqn\IIxib{
\ad {} , ~ \adp {} : \a  
\matrix{ {\rm homo.} \cr \longrightarrow\cr ~\cr  } 
\a ~ \phi (x) ~ \a . }
More precisely, 
\eqn\IIxii{
\ad{e_a} \( \ad{e_b} \( \phi (x) \) \) = 
\m abc \> \ad{e_c} \( \phi (x) \) . } 
This follows from the fact that the comultiplication $\De$ is 
a homomorphism from \am to $\a \ot \a$  \Ii{d} , and from 
\Ii{f} .  
For elements $a\in \a$ with the trivial comultiplication
$\De (a) = a\ot 1 + 1\ot a$, and counit $\vep (a) = 0$, 
$\ad{a}$ is just the usual commutator.

Repeated adjoint action on a given field $\phi (x)$ generates
a tower of its `descendents': 
\eqna\IIxiii
$$\eqalignno{
\phi (x) ~~ 
\matrix{ {\rm ad} \cr \longrightarrow \cr ~\cr }   
~~ 
\{ \phi_{a_1 ,a_2 ,...} (x) \}  &\equiv \Phib (x) 
&\IIxiii {a} \cr
\phi (x) ~~ 
\matrix{ {\rm ad'} \cr \longrightarrow \cr ~\cr }   
~~ 
\{ \phi_{a_1 ,a_2 ,...} ' (x) \}  &\equiv \Phib '(x) 
. &\IIxiii {b} \cr 
}$$
The form factors of any descendent of a field $\phi (x)$ are 
explicitly computable from knowledge of the form factors for 
$\phi (x) $ and the given action $\rh $ of \am on the space
of multiparticle states.  This is evident from the definition
\IIxi{}, where when computing the matrix elements of 
$\ad{e_a} \( \phi (x) \)$ the elements of \am to the right
or left of the field give a known transformation on states.  

By virtue of the symmetry of the S-matrix \IIv , the descendent 
form factors satisfy the axiom \IIvii .  
 However they do not 
generally satisfy \IIviii , hence descendents are generally not
local fields.  This is due to the non-locality of the 
conserved currents that generate \am.  Indeed, as explained 
in \rdenis\rbl , 
and similarly  for conformal field theories in 
\sigo ,
the non-trivial comultiplication of 
\am is due to the non-locality of these currents. 

A highest weight field is defined as a field which cannot be reached
by adjoint action on another field.  A precise definition of 
highest weight field requires a definition of raising and lowering
operators for the algebra \am , or at least for its quantum
double. The towers of descendents \IIxiii{} of a highest weight field
by construction fill out a representation 
\def\hw{\Lambda} 
\def\rhw{\rho_{\hw} } 
\def\rhwo#1{{\rho_{\hw_#1} }} 
$\rhw$ of \am , denoted $\Phib_{\hw} (x)$. This is due to the property
\IIxii\ of the adjoint action.  More specifically, 
\eqn\IIxiv{
\ad{e_a} \( \Phib_\hw (x) \) = \rhw (e_a ) \Phib_\hw (x) . } 
The fields are thus intertwiners for \am . 

Specific models are characterized by which representations 
$\rhw$ span its field content.  For algebras \am generated by
a finite number of elements these representations are expected to 
be finite dimensional.  For the case where \am is simply a finite
dimensional Lie algebra, this corresponds to the fact that fields are
irreducible tensors.  For infinite dimensional algebras \am the possibility 
arises that the representations $\rhw$ are infinite dimensional, 
Verma-module representations. 
For the example considered in the next section, this is the case.  

The fields $\Phib_\hw (x)$ are in part characterized by their
braiding relations.  We will prove the fundamental braiding relation
\eqn\IIxxi{
\Phib_{\hw_2} (y,t) \> \Phib_{\hw_1} (x,t) 
= \CR_{\rhwo 1 ,\rhwo 2 } \> \Phib_{\hw_1 } (x) \> \Phib_{\hw_2 } (y) 
~~~~~x<y, } 
where $\CR_{\rhwo 1 ,\rhwo 2 } $ is the universal $\CR$ matrix
specialized to the representations $\rhwo{{1,2}}$ of the fields
$\Phib_{\hw_{1,2}}$: 
\eqn\IIxvib{
\CR_{\rhwo 1 ,\rhwo 2} = \rhwo 1 \ot \rhwo 2 \( \CR \) .}  

One way to prove \IIxxi\ is to use a 
generalization of the locality theorem used in \rsmirii . One first  
uses this locality theorem to 
show that 
\eqn\IIxv{
\Phib_{\hw_1} (x,t) \> \Phib '_{\hw_2} 
(y,t) = \Phib ' _{\hw_2}  (y,t) \> \Phib_{\hw_1}  (x,t) 
~~~~x<y. } 
The proof of \IIxv\ is model dependent; however it is ultimately
due to the crossing symmetry of the S-matrix.  The relation 
\IIxv\ will be established for the example of the next section. 

The fields 
$\Phib_\hw (x)$ and $\Phib '_\hw (x)$ are related by the following
expression: 
\eqn\IIxvi{
\Phib ' _\hw (x) = \CR_{\rh , \rhw } ~ \Phib_\hw (x) , }
where 
$\CR_{\rh ,\rhw}$ is again the universal $\CR$-matrix evaluated
in the representations indicated.  The formula \IIxvi\ is 
established by showing that $\CR_{\rh , \rhw}$ must satisfy 
its defining relation \Iviii . 
To prove the last statement, note that 
\eqnn\IIxvii
$$\eqalignno{
\rh \ot \rhw \( \De (e_a) \) \Phib_\hw (x) 
&= \u ija \> e_i \> \ad{e_j} \( \Phib_\hw (x) \) \cr
&= \u ija \> \u klj \> e_i \, s(e_k ) \> \Phib_\hw (x) \> e_l \cr
&= \Phib_\hw (x) \> e_a , &\IIxvii \cr }$$
where in the last step we used the coassociativity \Iiii {c} , 
the defining properties of the antipode \Iiii {e} , and 
\Iiii {h} .  Similarly, 
\eqn\IIxviii{
\rh \ot \rhw \( \De ' (e_a ) \) \Phib ' _\hw (x) = \Phib ' _\hw (x) \> e_a . } 
Therefore  
$\Phib ' _\hw (x) e_a = \CR_{\rh ,\rhw} \Phib_\hw (x) e_a $ implies
\eqn\IIxix{
\rh \ot \rhw \( \De ' (e_a ) \) \CR_{\rh , \rhw } \Phib_\hw (x) 
= \CR_{\rh , \rhw } ~\( \rh \ot \rhw \) \( \De (e_a ) \) \Phib_\hw (x) , }
which establishes the relation \Iviii\ for $\CR_{\rh , \rhw}$. 

The relation \IIxvi\ can be inserted into \IIxv\ to derive the 
braiding relation \IIxxi\ for the fields.  
We have 
\eqnn\IIxx
$$\eqalignno{
\Phib_{\hw_2} (y) \> \Phib_{\hw_1 } (x) &= \CR^{-1}_{\rh ,\rhwo{2}} 
\Phib_{\hw_1} (x) \CR_{\rh ,\rhwo{2} } \> \Phib_{\hw_2} (y)  \cr
&= s(e_i ) \Phib_{\hw_1 } (x) e_j  ~ \rhwo{2} (e^i ) 
\rhwo{2} (e^j ) ~ \Phib_{\hw_2}
(y) \cr
&= \( \rhwo{1} (e_a )  \Phib_{\hw_1} (x) \) ~ \( \rhwo{2} (e^a ) 
 \Phib_{\hw_2} (y) \) , &\IIxx\cr }$$
where we have used \Ivii ,  and 
the homomorphism property \IIxii\  
for the relation \Iiv {a} in the dual space. 
From the expression \Ivii\ for $\CR$, one has thus established 
\IIxxi .  

The above  proof of the braiding relation \IIxxi\ relied 
on \IIxv , which is established in a specific model using form 
factors.  Eq. \IIxxi\ can be understood in an alternative 
model independent way.    We first show
that the adjoint action on a product of fields is given by the 
comultiplication, namely: 
\eqnn\IIxxii
$$\eqalignno{
\ad{e_a} \( \Phib_{\hw_2} (y) ~  \Phib_{\hw_1} (x) \) 
&= \u bca \> s(e_b ) ~  \Phib_{\hw_2} (y) ~  \Phib_{\hw_1} (x) ~ e_c \cr
&= \u bca \> \ad{e_b} \( \Phib_{\hw_2} (y) \) ~  \ad{e_c} 
\( \Phib_{\hw_1} (x) \) \cr
&= \rhwo{2} \ot \rhwo{1} \( \De (e_a ) \) \> \Phib_{\hw_2} (y) 
\> \Phib_{\hw_1} (x) .  &\IIxxii\cr } $$
Above we have used the coassociativity \Iiii {c} and 
\Iiii {h} .  Applying $\ad{e_a}$ to both sides of 
\IIxxi , and using \IIxxii , one finds that $\CR$ must 
satisfy the defining property of the universal $\CR$-matrix
\Iviii . 

Let us suppose that we can associate a conserved current 
$J^\mu _a (x)$, satisfying $\d_\mu J^\mu _a (x) =0$, 
to an element $e_a$ of \am , 
such that $e_a = \int dx J^t_a (x)$.  The braiding of these
currents with other fields can be expressed in a simple way. 
Note first that due to the (co)associativity \Iiii {b,c} 
of the algebra \am , there exists an adjoint representation
$\rho_{{\rm adj}} $ of \am , $\a^*$ whose matrix elements are given in 
terms of the structure constants: 
\eqna\IIxxiii
$$\eqalignno{ 
\bra b \rho_{{\rm adj}} (e_a ) \ket c &= \m acb  &\IIxxiii {a} \cr
\bra c \rho_{{\rm adj}} (e^a) \ket b &= \u abc &\IIxxiii {b} \cr
}$$
This fact is analogous to the fact that the structure constants 
of a Lie algebra form the adjoint representation of the 
algebra due to the Jacobi identity.  
The currents $J^\mu_a (x)$ are associated to the fields in 
$\Phib_{{\rm adj}} (x)$.  The braiding relation \IIxxi\ for 
$\rhwo{2} = \rho_{{\rm adj}}$ reads: 
\eqn\IIxxiv{
J_a^\mu (y) ~ \Phib_\hw (x) = \rhw \( R^b_a \) \Phib_\hw (x) ~ J^\mu_b (y) 
~~~~~x<y , } 
where 
\eqn\IIxxv{
R^b_a \equiv \u cba \> e_c . }
The above equations imply that the braiding with the 
currents can be simply determined from the comultiplication 
$\De$. 

\newsec{The example of Yangian Symmetry} 

In this section we apply the general theory of the last
section to models with Yangian symmetry.  As explained in
the introduction these models can be described by the action
\wzw .
For simplicity we consider only the case of $sl(2)$. 

\medskip\noindent 
4.1 ~~{\it Symmetries of the S-matrix} 

The spectrum of particles in this model is known to consist only
of kinks $\ket {\be ;\pm}$ transforming under the 2-dimensional 
spin-1/2 representation $\ket \pm$  of $sl(2)$.  The S-matrix has the 
following simple form 
\eqn\IIIii{
S_{12} (\be ) = s_0 (\be ) \( \be  - i\pi P_{12} \), }
where $P_{12}$ is the permutation operator: 
\eqn\IIIiii{
P_{12} = \inv{2} \( \sum_{a=1}^3 \sigma^a \ot \sigma^a + 1 \) , }
($\sigma^a $ are the Pauli spin matrices), and 
\eqn\IIIiv{
s_0 (\be ) = \frac{ \Gamma (1/2 + \be /2\pi i ) \> \Gamma (-\be /2\pi i ) } 
{\Gamma (1/2 - \be /2\pi i ) \> \Gamma (\be /2\pi i ) }  . } 

We now describe the symmetries of the S-matrix \IIIii . 
Let $\qo{a} , a= 1,2,3$ 
denote the global $sl(2)$ 
\def\sl{$sl(2)$} 
charges satisfying 
\eqnn\IIIv
$$\eqalignno{
\[ \qo a , \qo b \] &= f^{abc} \qo c  &\IIIv\cr
f^{abc} &= i\ep^{abc} . }$$
Define an additional charge $\qp a$, satisfying 
\eqn\IIIvi{
\[ \qo a , \qp b \] = f^{abc} \qp c . }
\def\f#1#2#3{f^{#1#2#3} } 
These charges have the following action on 1-particle states:
\eqn\IIIvii{
\qo a \> \ket \be = t^a \>
\ket \be , ~~~~~~~~\qp a \> \ket \be = t^a \be \> \ket \be , }
where $t^a = \sigma^a /2$, and $\ket \pm $ are 
eigenstates of $t^3$, $t^3 \> \ket \pm = \pm \ket \pm /2$.  
The action on multiparticle states is given by 
the comultiplication 
\eqna\IIIviii
$$\eqalignno{
\De (\qo a ) &= \qo a \ot 1 + 1\ot \qo a &\IIIviii {a} \cr
\De (\qp a ) &= \qp a \ot 1 + 1\ot \qp a + \al \> \f abc \> \qo b \ot \qo c , 
&\IIIviii {b} \cr }$$ 
where 
\eqn\IIIix{
\al = -\frac{2\pi i}{c_A} , } 
and $c_A = 2$ is the quadratic casimir in the adjoint representation 
\eqn\IIIx{
\f abc \f bcd = -c_A \delta^{ad} . } 

The comultiplication \IIIviii{} implies 
\eqn\IIIxi{
\qo a \> \ket {\be_n , \be_{n-1} , ..., \be_1 } = \sum_{i=1}^n 
t^a_i ~\ket {\be_n , \be_{n-1} , ..., \be_1 }   }
\eqnn\IIIxii
$$\eqalignno{
\qp a \> \ket {\be_n ,  ..., \be_1 } &= \( \sum_{i=1}^n 
t^a_i \, \be_i 
+ \al \f abc \sum_{i>j} t^b_i \> t^c_j \) 
\ket {\be_n , ..., \be_1 }   \cr
&\equiv T^a (\be_1 , \ldots , \be_n ) ~ \ket {\be_n , \ldots , \be_1} 
.&\IIIxii
\cr}$$
That these are symmetries of the S-matrix, i.e. that \IIv\ is satisfied, 
is easily checked by explicit computation.  The value \IIIix\ for 
$\al$ is a consequence of the crossing symmetry of $S_{12}$.  

The equations \IIIv , \IIIvi , and \IIIviii{}, are part of the defining
relations of the Yangian $\CY$, as formulated by Drinfel'd \rdrinb . 
The complete set of relations for the Yangian include the additional
`terrific' \foot{
We are here quoting Drinfel'd.} \  relations 
\eqna\IIIxiii
$$\eqalignno{
\[ \qp a , \[ \qp b , \qo c \] \] 
&- \[ \qo a , \[ \qp b , \qp c \] \] 
= \al^2 A^{abcdef} \{ \qo d , \qo e , \qo f \}  
&\IIIxiii{a}   \cr
\[ \[ \qp a , \qp b \] , \[ \qo c , \qp d \] \] 
 &+ \[ \[ \qp c , \qp d \] , \[ \qo a , \qp b \] \] 
 ~ 
 \cr &= \al^2 \( A^{abklmn} \f cdk + A^{cdklmn} {\f abk} \)  
 \{ \qo l , \qo m , \qp n \} ,  &\IIIxiii{b} \cr
 }$$
 where 
 $A^{abcdef} = \f adk \f bel \f cfm \f klm $, and 
 $\{ , \} $ denotes symmetrization
 \eqn\IIIxiiib{
 \{ x_1 , x_2 , ... , x_n \} = \inv {n!} \sum_{i_1 \neq i_2 \cdot
 \cdot\cdot \neq i_n } 
 x_{i_1} x_{i_2} \cdot\cdot\cdot x_{i_n }  .} 
  
 These last relations \IIIxiii{} were constructed to be consistent with 
 the comultiplication.  
 For $sl(2)$, eq. \IIIxiii{a} is superfluous, i.e. it follows from 
 \IIIvi\ with the right hand side of \IIIxiii{a} equal to zero. 
 For other groups, \IIIxiii{b} follows from \IIIvi\ and \IIIxiii{a}.

 The remaining Hopf algebra properties 
 of the Yangian are
 \eqn\IIIxiv{
 \vep (\qo a ) = \vep (\qp a ) = 0 , }
 \eqn\IIIxv{
 s(\qo a ) = - \qo a ;~~~~~~~~~s(\qp a ) = -\qp a + \al \f abc \qo b \qo c .} 
 The antipodes \IIIxv\ are derived from the definition 
 \Ii{e} . 

The Yangian is a Hopf algebra for any constant value of the 
parameter $\al $. Setting $\al =0$ one recovers half of the 
Kac-Moody algebra.  By `half' we mean the generators with 
either positive or negative eigenvalue of the derivation
element, or positive or negative frequency modes in the sense
of current algebra. The Yangian is the unique deformation of the Kac-Moody
algebra that preserves the finite dimensional Lie subalgebra \IIIv . 
It has a natural grading where $\qo a$ has degree zero, and $\qp a$ and 
$\al$ have degree $+1$.  Elements of the Yangian corresponding to 
the other half of the Kac-Moody algebra are missing in 
Drinfeld's work. 
For our purposes, more specifically for constructing the universal 
$\CR$-matrix that characterizes the braiding of fields, it is necessary
to introduce the quantum double of the Yangian $\CD (\CY )$. 
The deformation of the  other half of the Kac-Moody algebra is located 
in $\CD (\CY )$  as we will show in the next section. 

\bigskip\noindent
3.2 {\it The Yangian Double and Deformations of the Full Kac-Moody Algebra}

 \def\CY{{\cal Y}}
\def\sl{\widehat{sl}(2)}
\def\triangle{\Delta} 

As explained in the last section, 
for our goals an  extension of the Yangian $\CY$
is needed which we are now going to describe. For the sake of 
simplicity we consider only the $sl(2)$ case
\foot{Below we will use the special properties of the 
$sl(2)$ structure constants: $\f abe \f cde 
= \del ad \del bc -\del ac \del bd$.}.  For this case,  the second 
relation in \IIIxiii{} can be rewritten as follows 
\eqn\fedi{
\f acd [[Q^c_1,Q^d_1],Q^b_1]+f^{bcd}[[Q^c_1,Q^d_1],Q^a_1]
= -2 
\alpha ^2 \( \f acd \{ Q^b_0,Q^c_0,Q^d_1\} +f^{bcd}\{ Q^a_0,Q^c_0,Q^d_1\} \)
}
(by multiplying \IIIxiii{b} by $\f abi \f cdj$),  
while the first relation is satisfied trivially. As mentioned above, 
the Yangian
is a deformation of the universal enveloping algebra of the subalgebra
$\sl _+$ of the Kac-Moody algebra $\sl $. 
In this work, for reasons that will soon be apparent, 
we need only consider the $\sl $ loop algebra, which is
obtained from the Kac-Moody algebra by setting the level to zero.   
We begin by presenting several simple facts concerning the loop algebra itself.

The loop algebra $\sl $ has the  set of generators $J^a_n,
\ n \in \CZ$ satisfying the relations
\eqn\fedii{
[J^a_m,J^b_n]=f^{abc}J^c_{m+n} }
It contains two subalgebras $\sl _+$ and $\sl _-$ generated by
$\{ J^a_n,\> n\ge 0 \} $ 
and $\{ J^a_n,\> n<0 \} $ respectively. One can define the set
of generators in a more economical way; for example, $J^a_1,J^a_0,J^a_{-1}$
are sufficient to generate $\sl $, because the relations \fedii\ allow
one to construct the remaining  generators. However, one 
must impose additional relations due to the fact that  the free Lie algebra 
generated by $J^a_1,J^a_0,J^a_{-1}$ is larger than the loop algebra.
Using the relations \fedii\ one has many ways of defining $J^a_m ,~ |m|\ge 3$
in terms of
$J^a_0,J^a_{\pm 1}$.  
($J^a_{\pm 2}$ are defined uniquely for $\sl $.) 
Thus, it is necessary to require that these
different definitions define the same algebra. It can be shown that
for the $\sl $ case it is sufficient to require that two different
definitions of $J^a_3, ~ J^a_{-3}$ coincide:
\eqn\fediii
{f^{acd}[[J^c_m,J^d_m],J^b_m]+f^{bcd}[[J^c_m,J^d_m] ,J^a_m]=0
~~~~~
m=\pm 1. }
Provided \fediii\
is satisfied, all possible definitions of $J^a_m ,\   |m|\geq 3$
coincide automatically. The relations \fediii\ play the same role as the 
Serre relations \ref\kac{V. G. Kac, {\it Infinite Dimensional Lie 
Algebras,} Cambridge University Press, 1985.} . 
Thus the loop algebra $\sl $ is generated by
$J^a_{\pm 1},J^a_0$ satisfying \fedii , \fediii , 
with the  trivial comultiplication
\eqn\fediv{
\triangle \( J^a_m \) =J^a_m\otimes 1+1\otimes J^a_m ~~~~~~~m=0,\pm 1. }

The loop algebra can be equipped with the inner product
\eqn\fedv{
\langle J^a_m,J^b_n\rangle =\delta ^{ab}\delta _{m+n+1,0} .
}
The subalgebras $\sl _+,\sl _-$ are dual with respect to this 
inner product. So, the structure of the classical double can be introduced
\ref\rseme{M. Semenov-Tian-Shansky, Funct. Anal. Appl.  17 (1983) 259;
Publ. RIMS 21 (1985) 1237.} \
 and the classical $r$-matrix can be defined:
\eqn\fedvi{
r=\sum _{n\ge 0}J^a_n\otimes (J^a_n)^* =\sum _{n\ge 0}J^a_n\otimes J^a_{-n-1} .
}
For the tensor product of finite dimensional representations depending on
spectral parameters $\lambda ,\mu $, 
where $J^a_m\rightarrow \lambda ^m t^a,
 \mu ^m t^a$,  the $r$-matrix is equal to
\eqn\fedvii{
{1\over {\lambda -\mu}}t^a\otimes t^a
}
and coincides with  the celebrated classical $r$-matrix found by Sklyanin 
\ref\rskly{E. Sklyanin, `On Complete Integrability of the Landau-Lifshitz
equation', Leningrad preprint LOMI (1980).}.

Let us now return to the Yangian. We want to demonstrate first that $\CY$
is a deformation of the universal enveloping algebra
$U(\sl _+ )$ of $\sl _+$.
 The basis of $U(\sl _+)$ consists of the following
elements
\eqn\fedviii{
\{ J^{a_1}_{k_1}\cdots J^{a_n}_{k_n}\} \equiv J^A_K } 
where $k_i\ge 0, \{ \  \} $ means symmetrization, and $A,K$ are multi-indices. 
We define
\eqn\fedix{
\deg(J^A_K)=\sum_{k\in K}   k . }

Define the following elements of $\CY$:
\eqn\fedx{
Q^a_k=\inv{(-2)^{k-1}} 
f^{a{a_1}{a_2}}f^{{a_2}{a_3}{a_4}}\cdots f^{{a_{2k-4}}{a_{2k-3}}
{a_{2k-2}}}
\times [Q^{a_1}_1[Q^{a_3}_1\cdots [Q^{a_{2k-3}}_1,Q^{a_{2k-2}}_1]\cdots ] }
and consider the vector space whose basis consists of the elements
\eqn\fedxi{
\{ Q^{a_1}_{k_1}\cdots Q^{a_n}_{k_n}\} \equiv Q^A_K .  }
Using induction in the degree of the elements one can show
that the vector space defined in this way coincides with $\CY$, 
just  as the vector space generated by \fedviii\  coincides with
$U(\sl _+)$. An  important point is that the relations
\fedi\  allow one to express the difference between two possible 
definitions of $Q^a_3$ in terms of the elements of lower degree.
Thus $\CY$ is indeed a deformation of $U(\sl _+)$  where 
$Q^a_k$ corresponds to $J^a_k$. 
We present the following examples of the multiplication map in 
$\CY$: 
\eqnn\mult
$$\eqalignno{
Q^a_{0}Q^b_{0} &= Q^{a,b}_{0,0}+{1\over 2}f^{abc}Q^c_{0} &\mult\cr
Q^{a,b}_{0,0}Q^c_{0} &= Q^{a,b,c}_{0,0,0}+{1\over 2}f^{acd}Q^{b,d}_{0,0}
+{1\over 2}f^{bcd}Q^{a,d}_{0,0}\cr
&~~~-{1\over 12}(\delta ^{ac}\delta ^{bd}+\delta ^{bc}\delta ^{ad}
-2\delta ^{ab}\delta ^{cd})Q^d_{0}\cr
}$$
\eqn\fedxxv{
\Rightarrow ~~
m^{Q^a_0}_{{Q^b_0}{Q^c_0}}={1\over 2}f^{abc}, ~~~m^{Q^a_0}_{Q^{de}_{00}
{Q^c_0}}
={1\over {12}}(2\delta ^{de}\delta ^{ac}
-\delta ^{dc}\delta ^{ae}-\delta ^{ec}\delta ^{ad}) . }

For what follows we need the 
comultiplication of $Q^A_K$, which can be shown to be of the form
\eqn\fedxii{
\triangle (Q^A_K)= 
\sum_{\deg ^* Q^A_K=\deg ^* Q^B_L+\deg^* Q^C_M} 
\mu{\>}^{Q^B_L,Q^C_M}_{Q^A_K} \ Q^B_L\otimes Q^C_M } 
where 
\eqn\fedxiib{
  \deg^* \( Q^A_K \) =-\sum_{k\in K}(k+1) .}
For example,
\eqnn\fedxiii
$$\eqalignno{
\triangle \( Q^a_1 \) &=Q^a_1\otimes 1+1\otimes Q^a_1+\alpha f^{abc}Q^b_0\otimes
Q^c_0 &\fedxiii\cr
\triangle \( Q^a_2 \) &=Q^a_2\otimes 1+1\otimes Q^a_2
+\alpha f^{abc}(Q^b_0\otimes Q^c_1-Q^c_1\otimes Q^b_0)\cr
&~~~~~~-\alpha ^2(Q^b_0\otimes Q^{a,b}_{0,0}
+Q^{a,b}_{0,0}\otimes Q^b_0).\cr} $$ 

We wish to extend  $\CY$ in order to obtain a  deformation of the complete
loop algebra. 
This can be accomplished by 
following  the general construction of 
the quantum double discussed in Section 2. Namely, let us construct the Hopf
algebra $\CY^*$ dual to $\CY$. The element dual to $Q^a_0$ is denoted
by $2\alpha Q^a_{-1}$. It can be shown that: \hfil \break
\noindent
1. The algebra $\CY^*$ is generated by $Q^a_{-1}$.  \hfill\break
\noindent
2. The generators $Q^a_{-1}$ satisfy the relation: \hfill\break
\eqn\fedxiv{
f^{acd}[[Q^c_{-1},Q^d_{-1}], Q^b_{-1}] 
+ f^{bcd}[[Q^c_{-1},Q^d_{-1}],Q^a_{-1}] =0
}
\noindent
which is the undeformed form of \fediii\ for $m=-1$.
The relation \fedxiv\ will be proven below. 

The algebra $\CY^*$ is equivalent as an algebra
(but not as coalgebra!)
to $U(\widehat{sl}(2)_-)$.
So, its basis consists of
\eqn\fedxv{
Q^A_{-K} \equiv \{ Q^{a_1}_{-k_1}\cdots Q^{a_n}_{-k_n}\},
~~~k_i > 0
}
\noindent
where
\eqn\fedxvi{
Q^a_{-k}=\inv{(-2)^{k-1}}f^{a{a_1}{a_2}}f^{{a_2}{a_3}{a_4}}\cdots 
f^{a_{2k-1}a_{2k-3}a_{2k-2}} 
\times [Q^{a_1}_{-1}[Q^{a_3}_{-1}\cdots 
[Q^{a_{2k-3}}_{-1} , Q^{a_{2k-2}}_{-1}]\cdots ] .}
The degree of $Q_{\pm K}^A$ is defined as 
\eqn\degree{
\deg \( Q^A_{\pm K} \) = \sum_{k\in K} \pm k .} 
The multiplication map $m^*$  in $\CY^*$ is described by the equation:
\eqn\fedxvii{
Q^A_{-K}\> Q^B_{-L} =
\sum_{\deg Q^A_{-K}+\deg Q^B_{-L} =  \deg Q^C_{-M}}
{m^*}^{Q^C_{-M}}_{Q^A_{-K} , Q^B_{-L}}
\ Q^C_{-M}
.}
From the general theory of section 2, namely \Iiv{a}, the constants 
$m^*$ are computable from the comultiplication map in $\CY$.  

To exhibit the properties of the double $\CD (\CY )$, knowledge
of the basis elements is needed.  
The dual elements to $Q^A_K \in \CY$ should be expressed in terms
of the basis elements of $\CY^*$.  
Let $A,B,C \in \CY$ and $X,Y,Z \in \CY^*$.  The comultiplication
coefficients $\mu$ of $\CY$ and the multiplication map $m^*$ in
$\CY^*$ have the usual definitions:  $\Delta (A) = \mu^{BC}_A \> 
B\otimes C$ and $X\cdot Y = {m^*}^Z_{XY}$. 
Let us define a matrix
of coefficients $\CT$, where  
\eqn\dual{
A^* = \sum_{X\in \CY^*}  \, \CT^{AX} \> X ~~~~~~A\in \CY . } 
 Then by 
\Iiv{a}, $\CT$ must satisfy 
\eqn\tt{
\mu^{AB}_C \> \CT^{CZ} = \CT^{AX} \> \CT^{BY} \> {m^*}^Z_{XY} . }  
We were unable to deduce an general formula for $\CT$.  However 
a study of particular examples (up to degree-4 in $\CY$ ) convinced
us that the defining relations \tt\ allow a solution.  Furthermore
the matrix $\CT$ is triangular: 
\eqn\tri{
\CT^{AX} = 0 ~~~ {\rm if} ~~\deg^* (A) \neq \deg (X)~~~
{\rm or} ~~\deg (A) > \deg^* (X) , } 
where 
\eqn\degd{
\deg^* \( Q^A_{-K} \) = \sum_{k\in K} (k-1). } 

Let us present some examples of dual elements.  Consider
$(Q_1^a )^*$.  From \Iiv{a} one has
\eqn\dui{
\( Q_0^a \)^* \( Q_0^b \)^* = 
(2\al )^2 \( Q_{-1,-1}^{a,b} + \inv {2} \[ Q_{-1}^a , \qm b \] \) 
= \sum_{D\in \CY}  \mu_{D}^{\qo a , \qo b } \> D^* . } 
The only non-zero contributions to the above sum over $D$ comes from 
$\mu_{\qp c}^{\qo a , \qo b } = \al \f abc$ and 
$\mu_{Q_{00}^{ab}}^{\qo a , \qo b } = 1$.  Thus
\eqnn\duii
$$\eqalignno{ 
\al \f abc \( \qp c \)^*  
&= 2\al ^2 \[ \qm a , \qm b \] &\duii\cr   
\( Q_{00}^{ab} \)^* &= 4\al^2 \, Q_{-1,-1}^{a,b} . \cr}$$
The above equation implies $( \qp a )^* = 2\al Q_{-2}^a $ and 
$[ \qm a , \qm b ] = \f abc Q_{-2}^c $.  

One can deduce $(Q_k^a )^*$ inductively as follows.  Let us suppose
$( Q_{k-1}^a )^* = 2\al Q_{-k}^a $, and 
\eqn\duiii{
\Delta \( Q_{k-1}^a \) = Q^a_{k-1} \ot 1 + 1\ot Q_{k-1}^a 
+\al \f abc \( Q_0^b \ot Q_{k-2}^c - Q_{k-2}^c \ot \qo b \) + \cdots }
Note that $\Delta ( Q_2^a )$ is of this form.  
Representing $Q_k^a$ as $-\f abc \qp b Q_{k-1}^c$, 
from \duiii\ one shows that $\Delta (Q_k^a )$ is of the same form
as \duiii\ with $k-1 \to k$.  We have
\eqn\duiv{ 
\( \qo a \)^* \( Q_{k-1}^b \)^* = (2\al )^2 \qm a Q_{-k}^b 
= \sum_{D\in \CY} \mu_D^{\qo a , Q_{k-1}^b } \> D^* .} 
Now it is not hard to realize that the only contributions in the above
sum over $D$ come from $D=Q_k^a$ and $ Q_{0,k-1}^{ab}$.  Thus 
\eqnn\duv
$$\eqalignno{
\( Q_k^a \)^* &= 2\al \, Q_{-k-1}^a &\duv\cr 
\( Q_{0k}^{ab} \)^* &= 4\al^2 Q_{-1,-k-1}^{a,b} , \cr}$$
which implies 
\eqn\duvi{
\[ \qm a , Q_{-k}^b \] = \f abc Q_{-k-1}^c . } 
The `Serre' relation \fedxiv\ in $\CY^*$ is now seen to be a consequence
of \duvi . 
The relation  \duvi\ also implies the following examples of the 
multiplication map in $\CY^*$: 
\eqnn\fedxviii
$$\eqalignno{
Q^a_{-1}Q^b_{-1} &= Q^{a,b}_{-1,-1}+{1\over 2}f^{abc}Q^c_{-2} &\fedxviii\cr
Q^{a,b}_{-1,-1}Q^c_{-1} &= Q^{a,b,c}_{-1,-1,-1}+{1\over 2}f^{acd}Q^{b,d}_{-1,-2}
+{1\over 2}f^{bcd}Q^{a,d}_{-1,-2}\cr
&~~~-{1\over 12}(\delta ^{ac}\delta ^{bd}+\delta ^{bc}\delta ^{ad}
-2\delta ^{ab}\delta ^{cd})Q^d_{-3} . \cr
}$$

Dual elements generally have corrections to the classical formulas, unlike
\duv . Consider for example 
$\( Q_{000}^{abc} \)^*$.  By examining 
\tt\ with $A= Q_{00}^{ab}$ and $B= Q_0^c$, one finds 
\eqn\duvii{
\( Q_{000}^{abc} \)^* = 8\al^3 \, Q_{-1,-1,-1}^{a,b,c} 
+ \frac{4}{3} \al^3 
\( \del ac \del bd + \del bc \del ad 
+ \del ab \del cd \) \, Q_{-3}^d . }

Now using the general formulas \Ix{}  one can deduce the 
commutation relations in the Yangian double.  A straightforward
computation gives 
\eqnn\IIIxixp
$$\eqalignno{
\[ \qo a , B^* \] &= m^B_{[D,\qo a ]} D^* &\IIIxixp\cr
\[ \qp a , B^* \] &= 
m^B_{[D,\qp a ]} \> D^* 
+ \al \f acd \( m^B_{\qo c \, D} ~+~ m^B_{D\, \qo c } \) D^* \qo d 
\cr 
&~~~~~~~+\al \f acd m^B_{\qo c , [D,\qo d ]} \> D^* 
\cr}$$
where $B,D \in \CY $, 
$m^B_{[D,\qo a ] } = m^B_{D\, \qo a} - m^B_{\qo a \, D}$, 
and similarly for  
$m^B_{\qo c ,[D,\qo d ]}$.  
From \IIIxixp\ with $B= \qo b$, and the above dual elements, one finds 
\eqn\IIIxxi{
\[ \qo a , \qm b \] = \f abc \qm c .}
One can also compute the 
first few terms in the commutator 
\eqnn\IIIxxii
$$\eqalignno{ 
\[ \qp a , \qm b \] &= \f abc \qo c  - \frac{\al}{2} \f abc \qm c 
- 2 \al^2 Q^{a,b}_{-1,-1}  + \frac{2 \al^2}{3} \de^{ab} \>  Q^{c,c}_{-1,-1} 
\cr
&+\frac{2 \al^2}{3} \f abd Q^{c,c}_{-1,-1}  \qo d  - \frac{2 \al^2}{3} 
\f acd Q^{c,b}_{-1,-1}  \qo d + \ldots 
&\IIIxxii\cr }$$

The comultiplication of $Q^a_{-1}$ can be found as follows. Consider
all possible products of elements of $\CY$ which contain $Q^a_0$.
Evidently the only elements of this type are $Q^{a_1\cdots a_n}_{0,\cdots 0}$:
\eqn\fedxxiv{
Q^{{b_1}\cdots {b_k}}_{{0}\cdots 0}Q^{{c_1}\cdots {c_l}}_{0\cdots 0}=
m^{Q^a_0}_{Q_{0\cdots 0}^{{b_1}\cdots {b_k}},
Q^{{c_1}\cdots {c_l}}_{0\cdots 0}}\cdot Q^a_0  + \cdots }
According to the general rules
\eqn\fedxxvi{
\Delta \( Q^a_{-1} \) =\inv{2\alpha} \sum _{C,D\in \CY}m^{Q^a_0}_{DC} ~ 
C^* \otimes D^*  .} 
The first 
several terms in \fedxxvi\  are
\eqnn\IIIxxiii
$$\eqalignno{ 
\De \( \qm a \) &= \qm a \ot 1 + 1\ot \qm a 
-\al \f abc \qm b \ot \qm c 
+\frac{2 \al^2}{3} 
\(\del ac \del bd + \del ba \del dc -2 \del bc \del ad \) 
\cr
&~~~~\cdot\( \qm d \ot Q^{b,c}_{-1,-1} + Q^{b,c}_{-1,-1}  \ot \qm d 
\) + \ldots &\IIIxxiii\cr } $$

Finally the $\CR$-matrix of ${\cal D}(\CY )$ is equal to 
\eqnn\fedxxviii
$$\eqalignno{
\CR=I+
& 2\al \sum_{k=0}^\infty Q_k^a \ot Q_{-k-1}^a 
~+~ 4\al^2 Q_{0k}^{ab} \otimes Q_{-1,-k-1}^{ab}  \cr 
& +8 \al^3 Q_{000}^{abc} \otimes Q_{-1,-1,-1}^{a,b,c} 
+ 4\al^3 Q_{000}^{aab} \otimes Q_{-3}^b 
+\cdots
&\fedxxviii\cr }$$
In the classical limit
\eqnn\clasr
$$\eqalignno{
R &= 1+\alpha r+o(\alpha^2 ) &\clasr\cr
r &= Q^a_0\otimes 
Q^a_{-1}+Q^a_1\otimes Q^a_{-2}+Q^a_2\otimes Q^a_{-3}+\cdots \cr
}$$
which coincides with the classical $r$-matrix \fedvi .

We remark that one does not have the freedom to introduce a level $k$ into
the Yangian double, as in \km .  Indeed, upon setting $\alpha$ to zero in 
\IIIxxii , one recovers the loop algebra.  Though it remains an interesting
mathematical problem to determine whether a central extension of the
Yangian can be introduced by other means, this does not appear to be
relevant for the physical applications we are considering.

\def\ad#1{ {\rm ad}_{#1} } 
\def\adp#1{ {\rm ad'}_{#1} } 

\bigskip\noindent 
3.3 ~{\it Adjoint Action on Fields} 

From the definition \IIxi{}  we have the following adjoint
actions 
\eqna\IIIxvi
$$\eqalignno{
\ad{\qo a } \( \phi (x) \) &= \adp{\qo a} \( \phi (x) \) 
= \[ \phi (x), \qo a \] &\IIIxvi{a} \cr
\ad{\qp a} \( \phi (x) \) &= \[ \phi (x) , \qp a \] 
- \al \> \f abc \> \qo b \> \[ \phi (x) , \qo c \] 
&\IIIxvi{b} \cr
\adp{\qp a} \( \phi (x) \) &= \ad{\qp a } \( \phi (x) \) 
+ 2\al \> \f abc \> \qo b \> \[ \phi (x) , \qo c \] . &\IIIxvi{c} \cr}$$

Highest weight fields $\phi_\hw (x)$ can be defined to satisfy
\eqn\IIIxvii{
\ad{\qo + } \( \phi_\hw (x) \) = \ad{\qm a} \( \phi_\hw (x) \) =0 , } 
where $Q_0^\pm = \( \qo 1 \pm i \qo 2 \) /\sqrt{2} $. 
The vacuum is invariant under all the charges 
\eqn\IIIxviii{
\qo a \rvac = \qp a \rvac = \qm a \rvac = 0 . } 
Descendents of $\phi_\hw (x)$, which comprise the multiplet 
$\Phib_\hw (x)$, are denoted as 
\eqn\IIIxix{
{\phi_\hw}^{a_1 ,\cdots , a_m ; N} (x) = 
\ad{\qp {a_1}}  \, \circ \, \ad{\qp {a_2 } }\, \circ \cdots \circ 
\ad{\qp {a_m} } \, \circ \,  
\ad{\qo - } ^N ~ \( \phi_\hw (x) \) . } 

Let $f(\be_1 , \ldots , \be_n )$ denote the form factor
for a field $\phi (x)$, not necessarily
highest weight, as in \IIvi . We suppose that $\phi (x)$
transforms under some finite dimensional representation 
of $sl(2)$: 
\eqn\IIIxx{
\[ \qo a , \phi (x) \] = t^a_\phi \> \phi (x) . } 
The $sl(2)$ invariance implies 
\eqn\IIIxxi{
\( t^a_\phi + \sum_{i=1}^n t^a_i \) \> f(\be_1 , \ldots , \be_n ) =0. } 
Consider now the form factors of the descendents with respect to 
$\qp a$
\eqn\IIIxxii{
f^{a_1 ,a_2 ,\ldots , a_m} (\be_1 , \ldots , \be_n ) 
= \lvac \> \phi^{a_1 , a_2 , \ldots , a_m ; 0} (0) \> 
\ket {\be_n , \ldots , \be_1 } . } 
From \IIIxvi{}, \IIIxviii , and \IIIxii , one obtains these form 
factors in terms of the form factors of $\phi (x)$
\eqn\IIIxxiii{
f^{a_1 , a_2 , \ldots , a_m } ( \be_1 , \ldots , \be_n ) 
= T^{a_1} (\be_1 ,\ldots , \be_n ) 
\cdots  T^{a_m} (\be_1 ,\ldots , \be_n ) 
~f(\be_1 , \ldots , \be_n ) . } 
A similar expression holds for descendents with respect to
$\adp{\qp a } $, where $\al \to -\al $ in the definition for 
$T^a$.  

We now show that the two sets of $\phi$'s descendent fields
$\Phib$ and $\Phib ' $ commute for $x<y$ \IIxv . 
The proof of \IIxv\ relies on a generalization of the locality
theorem, which can be stated as follows \rsmiri .  
Let the form factors of two fields $\CO_1 (x)$ and $\CO_2 (x)$ be 
given.  The matrix elements are defined as 
\eqn\IIIxxiv{
f_{1,2} (\al_m ,\ldots , \al_1 | \be_1 , \ldots , \be_n ) 
= \bra{ \al_1 , \ldots , \al_m } \> \CO_{1,2} (0) \> \ket{ \be_n , \ldots , 
\be_1} 
.}
These matrix elements can be defined in two different ways
\eqna\IIIxxv
$$\eqalignno{
f_{1,2} (\al_m ,\ldots , \al_1 | \be_1 , \ldots , \be_n ) 
&= 
f_{1,2} (\al_m -i\pi  ,\ldots , \al_1 -i\pi , \be_1 , \ldots , \be_n ) 
&\IIIxxv{a}\cr
f '_{1,2} (\al_m ,\ldots , \al_1 | \be_1 , \ldots , \be_n ) 
&= f_{1,2} (\be_1 ,\ldots , \be_n ,   \al_m +i\pi  , \ldots , \al_1 +i\pi ) 
&\IIIxxv{b} \cr }$$
(For simplicity of notation we don't display the charge conjugation
matrices.) 
For local operators, by \IIviii , $f=f'$.  The locality theorem 
states that all matrix elements satisfy 
\eqn\IIIxxvi{
\CO_1 (x) \> \CO_2 (y) = \CO_2 (y) \> \CO_1 (x) ~~~~~x<y . } 
The proof of \IIIxxvi\ involves inserting a complete set of states
between the operators, representing the matrix elements of $\CO_1$ 
and $\CO_2$ with $f_1$ and $f'_2$ respectively, and using 
analytic properties of form factors \rsmiri . 

The details of the proof of the locality theorem admit the
following generalization \rsmirii .  Let us be given any two operators 
$\CO (x)$ and $\CO ' (y)$, whose matrix elements are of 
the form $f$ and $f'$ respectively, as in \IIIxxv{}.  Then 
\eqn\IIIxxvii{
\CO (x) \> \CO '(y) = \CO ' (y) \> \CO (x) ~~~~~x<y . } 

Let us apply the above result to the comparison of descendent 
form factors in 
$\Phib$ and $\Phib '$.  
For simplicity consider the case where the ancestor field 
$\phi$ is local.  
We adopt the notation that $\bra A , \ket B$ represent the states 
$\bra {\al_1 , \ldots , \al_m }$ and 
$\ket { \be_n , \ldots , \be_1 } $, respectively.  
The form factor $f^a (\be_1 , \ldots , \be_n )$ has the 
following analytic property 
\eqn\IIIxxviii{
f^a (\be_1 , \ldots , \be_n + 2\pi i ) = T^a (\be_1 , \ldots , \be_n ) 
\> f(\be_n , \be_1 , \ldots , \be_{n-1} ) 
+ 2\pi i \> t^a_n \, f(\be_n , \be_1 , \ldots , \be_{n-1} ) .} 
The operator $T^a$ has the following property
\eqn\IIIxxix{
T^a (\be_1 ,\ldots , \be_n ) = T^a (\be_n , \be_1 , \ldots , \be_{n-1} ) 
-2\pi i \, t^a_n \> -2\al \f abc \, t^b_B \> t^c_n , } 
where $t^b_B$ is the representation of $sl(2)$ on the state 
$\ket{B}$, $t^b_B = \sum_{i=1}^n t^b_i $. 
Using \IIIxxi , the equations \IIIxxviii\ and \IIIxxix\ imply 
more generally that 
\eqn\IIIxxx{
f^a (B_1 , B_2 + 2\pi i ) = f^a (B_2 , B_1 ) 
+ 2\al \f abc \> t^b_\phi \> t^c_{B_2} \> f(B_2 , B_1 ) . } 

Consider now the two kinds of matrix elements 
\eqna\IIIxxxi
$$\eqalignno{
f^a (A|B) &= f^a (A-i\pi , B) &\IIIxxxi{a} \cr
{f'}^a (A|B) &= f^a (B , A+ i\pi ) . &\IIIxxxi{b} \cr }$$
From \IIIxxx 
\eqn\IIIxxxii{
{f'}^a (A|B) = f^a (A-i\pi , B ) + 2\al \f abc \> t^b_\phi \> t^c_A \> 
f(A-i\pi , B ) . } 
Comparing \IIIxxxii\ with \IIIxvi{}, one sees that 
$f^a (A|B) $ and $f'^a (A|B) $ represent the matrix elements of 
$\ad{\qp a} \( \phi \) $ and $\adp{\qp a} \( \phi \) $ 
respectively.   Therefore, from the generalized locality 
theorem, these two fields commute for $x<y$.  Repeated adjoint
action yields the result \IIxv . 

Let $J^a_\mu (x)$ and $\CJ^a_\mu (x)$ denote the currents generating
the conserved charges $\qo a$ and $\qp a$ respectively.  
Form the general result \IIxxiv\ one has the following 
braiding relations 
\eqnn\IIIxxxiib
$$\eqalignno{
J^a_\mu (y) \> \Phib_\hw (x) &= \Phib_\hw (x) \> J^a_\mu (y)  ~~~~~x<y 
&\IIIxxxiib\cr
\CJ^a_\mu (y) \> \Phib_\hw (x) &= \Phib_\hw (x) \> 
\CJ^a_\mu (y) 
-\al \f abc \[ \qo b , \Phib_\hw (x) \] \> J^c_\mu (y) . \cr}$$
In applying \IIxxiv\ we have used the fact that the current 
generating the identity is zero. 

Non-local currents $\CJ^a_\mu (x)$ generating the charges $\qp a$ 
were  constructed by L\"uscher and Bernard \rlus\rdenis . 
As was shown in these works, the current $\CJ^a_\mu (x)$ can be
constructed solely in terms of the current $J^a_\mu (x)$, through
a non-local bilinear expression.  
Using this explicit construction of the current it was shown in 
\rdenis\ that it satisfies the braiding relation \IIIxxxiib . 

Let us say a few words about what one expects for the field 
content of the specific model we are considering.  The conserved 
energy momentum tensor $T_{\mu\nu}$, being an $sl(2)$ singlet, 
of course has no descendents with respect to $\qo a$.  However
repeated $\ad{\qp a}$ action generates an infinite multiplet 
of conserved currents, which we analyze in more detail in the 
next section.  As we will show, the energy momentum tensor
and the $sl(2)$ current $J^a_\mu$ are in the same multiplet. 
There are also fields $\psi_{\pm} (x) , \bar{\psi}_{\pm} (x)$ 
which are $sl(2)$ doublets that create the kink particles 
asymptotically.  
The fields $\psi_+ , \bar{\psi}_+$ are highest weight fields for
some additional infinite dimensional Verma module representations. 

\bigskip\noindent
3.4 ~~{\it Descendents of the Energy-Momentum Tensor}

The form factors of the energy-momentum tensor $T_{\mu\nu} (x)$ 
and the $sl(2)$ current $J^a_\mu (x)$ are known 
\rsmiri .  They have the following presentation.  Define
two operators $\CO_T$ and $\CO_J^a $, related to 
$T_{\mu\nu}$ and $J^a_\mu$ by 
\eqnn\IIIxxxiii
$$\eqalignno{
T_{\mu\nu} (x) &= \inv{2} \( \ep_{\mu\mu '} \ep_{\nu\nu '} 
+ \ep_{\nu\mu '} \ep_{\mu\nu '} \) \> \d_{\mu '} \d_{\nu '} \> 
\CO_T (x)  \cr 
J^a_\mu (x) &= \ep_{\mu\nu} \d_{\nu} \CO^a_J (x) , &\IIIxxxiii
\cr}$$
where $\ep_{\mu\nu}$ is the antisymmetric tensor.  
The energy-momentum tensor and $sl(2)$ current can be represented
this way since they are conserved.  The form factors for 
$T_{\mu\nu}$ and $J^a_\mu$ follow simply from those of 
$\CO_T , \CO^a_J$ due to the relation 
\eqn\IIIxxxiv{
\bra A \> \d_\mu \CO (x) \> \ket B = 
i \( \sum_{i=1}^m p_\mu (\al_i ) - \sum_{i=1}^n p_\mu (\be_i ) \) 
\bra A \> \CO (x) \> \ket B . } 

For simplicity we discuss only the two-particle form factors. 
They take the following form 
\eqnn\IIIxxxv
$$\eqalignno{
\lvac \> \CO_T (0) \> \ket {\be_2 , \be_1 } &= 
\inv{2\pi i} \> \frac{\zeta (\be_1 -\be_2 ) }{\be_2 -\be_1 -i\pi} 
~ \ket {{\rm singlet}} _{21}  \cr
\lvac \> \CO^a_J (0) \> \ket {\be_2 , \be_1 } &= 
\frac{\zeta (\be_1 -\be_2 ) }{2\pi i } \> 
~ \ket {{\rm triplet};a} _{21} ,  
&\IIIxxxv\cr }$$
where $\zeta (\be )$ is the function
\eqn\IIIxxxvi{
\zeta (\be ) = \sinh (\be /2 ) \> \exp 
\( \int_0^\infty \, dx \> \frac{\sin^2 \( x(\be + i\pi )/2\pi \)}
{x\sinh (x) \cosh (x/2 )}  
e^{-x/2}  \). } 
The states in \IIIxxxv\ are 
\eqnn\IIIxxxvii
$$\eqalignno{
\ket { {\rm singlet} } _{21} &= \( \ket + \ot \ket - \> -\> 
\ket - \ot \ket + \) &\IIIxxxvii\cr
\ket { {\rm triplet};  1,2,3  }_{21} 
&=  \( \ket - \ot \ket -  - \ket + \ot \ket + \) /2 , 
~~~i \( \ket - \ot \ket - + \ket + \ot \ket + \) /2 , 
\cr
&~~~~ \( \ket + \ot \ket - + \ket - \ot \ket + \) /2  . 
\cr }$$

Consider now the first descendent of $T_{\mu\nu}$ whose form factors
are given by the general formula \IIIxxii .  Since 
$T_{\mu\nu}$ is an $sl(2)$ singlet, its first descendent is
a local operator, as indicated by \IIIxxx . One can show by a simple
computation that 
\eqn\IIIxxxviii{
T^a (\be_1 , \be_2 ) ~\ket { {\rm singlet} } _{21} 
= \( \be_2 - \be_1 - i\pi \) ~\ket { {\rm triplet} ; a} 
_{21} . } 
Therefore, comparing the 2-particle form factors of this 
first descendent of $T_{\mu\nu}$ with the form factors of 
$J^a_\mu$, one derives 
\eqn\IIIxxxix{
\[ \qp a , T_{\mu\nu} (x) \] = -\inv{2} 
\( \ep_{\mu\al } \d_{\al} J^a_\nu (x) 
+ \ep_{\nu\al} \d_\al J^a_\mu (x) \) . } 
Introduce the generator $L$ of Lorentz boosts 
\eqn\IIIxxxx{
L= -\int dx \> x T_{00}  ~~~~~(t=0) . } 
Then the global version of \IIIxxxix\ is 
\eqn\IIIxxxxi{
\[ L , \qp a \] = - \qo a . } 
The generator $L$ thus has degree $-1$. 
The equation \IIIxxxxi\ was expected from the fact that 
on-shell, one has \IIIvii\  and $L= -\der{\be}$.  
Thus one concludes that the Yangian must actually be 
extended to include the Poincar\'e algebra with generators
$L, P_\mu$ in order to realize its full implications. 
Though we have presented the derivation of 
\IIIxxxix\ using only the two particle form factors, 
the same relation is implied by the general 
multiparticle form factors. 

We consider next the first descendents of $J^a_\mu$.  
One finds 
\eqn\IIIxxxxii{
\inv{2} \f abc ~\ad{\qp a } \( J^b_\mu  (x) \) 
= \CJ^c_\mu (x) + \frac{\al}{2} \, J^c_\mu (x) ,  } 
up to possible additional total derivatives.  This follows by
comparing the braiding of the current $\CJ^a$ on the right hand
side of \IIIxxxxii\ computed from the generalized locality
theorem with the braiding of the more abstractly defined current 
in \IIIxxxiib , and also by considering the global relation \IIIvi . 

\def\jt{J} 
\def\ot{\CO} 
\def\qt{Q} 

As a final example consider the conserved current and charge 
\eqn\IIIxxxxiii{
\CN_\mu (x) = \ad{\qp a} \( J^a_\mu (x) \) \, ; ~~~~~~~~
N = \int dx ~ \CN_t (x) . } 
In order to characterize this current, we determine the matrix
elements of the charge $N$.  Matrix elements of conserved charges
can be generally computed as follows. 
Let $\jt_\mu (x)$ be a general current for a conserved charge 
$\qt$, and define $\jt_\mu (x) = \ep_{\mu\nu} \d_\nu \ot (x)$. 
The operator $\ot$ is non-local.  The form factors of 
$\ot (x)$ correspond to the partially integrated current 
\eqna\IIIxxxxiv
$$\eqalignno{
\int_{-\infty}^x dy ~ \bra A \> \jt_t (y) \> \ket B 
&= \bra A \> \ot (x) \> \ket B  &\IIIxxxxiv{a} \cr
\int_{x}^\infty  dy ~ \bra A \> \jt_t (y) \> \ket B 
&= - \bra A \> \ot (x) \> \ket B  .&\IIIxxxxiv{b} \cr
}$$
The matrix elements on the RHS's of \IIIxxxxiv{a,b} 
should be defined by the continuations in \IIIxxv{a,b} 
respectively, since these are consistent with the range of 
integration. 
Therefore
\eqn\IIIxxxxv{
\bra A \> \qt \> \ket B = \lim_{\ep \to 0} ~~
f (A+i\ep |B ) - f (A-i\ep |B) , }
where $f$ are the form factors of $\ot$.  The 
RHS of \IIIxxxxv\ is non-zero due to the fact that 
$\ot$ is non-local. 

Returning to the charge $N$, let $\CN_\mu (x) = 
\ep_{\mu\nu} \d_\nu \CO_\CN (x)$.  The form factors of 
$\CO_\CN $ are computed to be 
\eqn\IIIxxxxvi{
\lvac \> \CO_\CN (0) \> \ket {\be_2 , \be_1 } 
= - \frac{c_{1/2} }{2\pi i} \> \zeta (\be_1 - \be_2 ) 
\( \be_1 - \be_2 -i\pi \) \ket {{\rm singlet}} , } 
where $c_{1/2} = 3/4$ is the casimir in the spin 1/2 representation. 
Using the fact that $\zeta (\be_1 -\be_2 )$ has a simple pole 
at $\be_2 = \be_1 + i\pi$, and the identity 
\eqn\IIIxxxxvii{
\inv{ \be - i\ep } - \inv{ \be + i\ep } = 2\pi i \, \delta (\be) , 
} 
we obtain
\eqn\boop{
\bra {\be_1 , \mp } \> N \> \ket {\be_2 , \pm } = \frac{3\pi i}{2}  
\, \delta (\be_1 - \be_2 ) . } 
Thus on one-particle states, $N$ is just proportional to the particle
number operator.  A similar computation for multiparticle states
verifies this identification: 
\eqn\IIIxxxxix{
N\, \ket {\be_n , \ldots , \be_1 } = \frac{3\pi i}{2}  ~ n 
\> \ket {\be_n , \ldots , \be_1 } . } 

\bigskip
\newsec{Concluding Remarks} 

The relation \IIIxxxix\ expressing the derivative of the 
$sl(2)$ current as a descendent of the energy-momentum tensor
is remarkable.  It implies the space-time properties of the
theory are closely related to the $sl(2)$ properties.  
Since the non-local current generating $\qp a$ can be expressed
in terms of the $sl(2)$ current $J^a_\mu (x)$, 
\IIIxxxix\ implies the energy-momentum tensor can be 
expressed exactly in terms of the $sl(2)$ current also, 
thereby indicating an exact Sugawara construction in
the massive theory.  The details of such a construction
deserves further study.  This idea taken together with the 
fact that the action \wzw\ is expressed only in terms of 
the current $J^a_\mu $ implies that the theory is fully
characterized by the current algebra and the Yangian symmetry. 

We have shown that the level $k$ of the Kac-Moody symmetry in the 
conformal model does not enter into the Yangian algebra.  This is
somewhat paradoxical, since the spectrum of fields in the 
WZWN model is governed by $k$.  Furthermore, the level $k$ is still
defined in the massive theory as appearing in the current 
commutator
\eqn\curr{
\[ J^a_t (x) , J^b_x (0) \] = 
\f abc J^c_x (0) \, \delta (x) - \frac{k}{i\pi} \> \del ab 
\, \delta ' (x) . } 
The resolution of this puzzle can be found by considering the 
exact spectrum and S-matrices of the model \wzw\ for arbitrary
level $k$, which were proposed in \ref\rabl{\ABL} .  Similar
S-matrices were found for spin-chain realizations of \wzw\ in 
\ref\resh{N. Yu. Reshetikhin, {\it S-Matrices in Integrable Models of
Isotropical Magnetic Chains}, Harvard preprint HUTMP-90-B292, 1990.} . 
Here it was found that the spectrum of massive particles 
$K^\pm_{ab}$, $a,b \in \{0,1/2,1,\ldots , k/2 \}$ is still
governed by an integer $k$, and reflects the 
spectrum of primary fields of the conformal model $S_G$ it is 
a perturbation of.  However the S-matrix factorizes into
two pieces, one which is the  Yangian invariant factor 
\IIIii\ for the indices $\pm$ of $K^\pm_{ab}$, the other is 
characterized by an additional fractional supersymmetry of order $k$ 
which acts on the indices $a,b$.  Thus the Yangian symmetry is 
unaffected by varying the level $k$.  Rather, increasing $k$ enlarges the
Yangian symmetry to include an independent fractional supersymmetry. 

The precise mathematical nature of the field representations of the 
Yangian should be explored.  The mathematics literature has so far 
focused only on finite dimensional rapidity dependent representations, 
which are only applicable to on-shell objects.  Indeed this 
partly explains why the precise connection of the Yangian with 
a deformation of the full Kac-Moody algebra, which arises is the study
of its quantum double, was never precisely made before.  

The correlation functions are also constrained by the quantum 
symmetry, and we conclude with a few remarks on this problem.  It is evident 
that the Ward identities are given by 
\eqn\ward{
\De^{(n)} \( e_a \) ~ \lvac 
~ \phi_1 (x_1 ) ~ \cdots ~ \phi_n (x_n ) ~ \rvac  
 = 0. } 
 This is derived by inserting the generator $e_a$ to the left of the fields
 in the correlation function and using 
 $\lvac \> e_a = 0$, with \IIxxii . 
 This equation can be used to relate correlation functions of various
 fields related by adjoint action.  However by itself, it cannot 
 determine correlation functions.  What is missing are the 
 analogs of `null-fields' 
 that are available in the conformal theory. 
 The possibility of such null-fields is a very interesting issue. 

\bigskip
\centerline{Acknowledgements} 
\medskip
We gratefully thank D. Bernard, I. Frenkel,   N. Reshetikhin,
and M. Semenov-Tian-Shansky for valuable 
discussions. 
We wish to thank J. Cardy for the opportunity to visit the Institute
in Santa Barbara, where this work was begun.  F.S. thanks the 
Cornell group for hospitality and the MSI at Cornell for financial
support.   This work was supported in part by the US National Science 
Foundation under grants no. PHY-8715272 , PHY-8904035, and by
the US Army.

\listrefs
\bye